\documentstyle[twocolumn,prc,aps,epsf,epsfig]{revtex}
\def\<{\langle}
\def\>{\rangle}
\def\d{\partial}
\def\+{\dagger}

\def\U1A{U(1)$_{\rm A}$}

\begin{document}
 
\newcommand{\be}{\begin{eqnarray}}
\newcommand{\ee}{\end{eqnarray}}
\twocolumn[\hsize\textwidth\columnwidth\hsize\csname@twocolumnfalse\endcsname
\title{
 Domain Wall Bubbles \\ in High Energy Heavy Ion Collisions 
}

\author{ E.V. Shuryak$^1$ and A.R.Zhitnitsky$^2$}
\address{$^1$ 
Department of Physics and Astronomy, State University of New York, 
Stony Brook NY 11794-3800, USA}
\address{$^2$ Department of Physics and Astronomy, University of
British Columbia, Vancouver, BC V6T 1Z1, Canada}

\date{\today}
\maketitle

\begin{abstract}

It has been recently shown that meta-stable domain walls  
exist in  high-density QCD ($\mu\neq 0$)
 as well as in QCD with large number of colors
($N_c\rightarrow\infty$), with
the lifetime being exponentially long in both cases. 
Such metastable domain walls
 may exist in our world as well, especially in hot hadronic matter
with temperature close to critical.
In this paper we discuss what happens if a bubble made of such wall
is created in heavy ion collisions, in the mixed phase
between QGP and hadronic matter. We show it will further be expanded to larger
volume $\sim 20 fm^3$ by the pion pressure, before it  disappears,
 either by puncture or
contraction. Both scenarios leave distinctive 
experimental signatures of such events,
 negatively affecting the interference correlations between the
outgoing pions. 
\end{abstract}
\vspace{0.1in}
]
\begin{narrowtext}
\newpage

\section{ Introduction}
 Domain walls are common  in field theory,  
the simplest in the family of topological objects. If  they are
configurations of fields interpolating between two distinct vacua,
the domain wall obviously cannot decay. However even in
theories with  a $single$ vacuum   
domain-wall configurations may  nevertheless exist. 
Such domain walls are solitons which usually require a   discrete-type
of symmetry. In QCD such a symmetry indeed exists  and corresponds to the 
discrete rotations of the
so-called  $\theta$ angle $\theta\rightarrow\theta+2\pi n$.
 
The $\theta$ parameter  is not a dynamical field in QCD\footnote{We do not
discuss axions in this paper.}, and therefore it can not make 
a domain wall  by itself.  However, the $\theta$ angle appears in the
low energy description of QCD together with the physical $\eta' (x)$ 
field in a very special combination $(\theta- i\log Det U)$ where
the unitary matrix $U$ describes the pseudo-goldstone fields and the 
singlet $\eta'(x)$ field. Therefore, the domain walls which  may exist due to 
the discrete symmetry discussed above, can be realized by the 
physical $\eta'(x)$ field. If the $\eta'$ field is the light field,
all  calculations are under theoretical control and one can argue
that   the life time of the domain wall is parametrically large
in large $N_c$ limit  $\tau_{life} \sim \exp(  N_c^2)$,
\cite{fz}, as well as in large $\mu$ limit $\tau_{life} \sim \exp( \mu^{2+b}),
b=\frac{11}{3}N_c-\frac{2}{3}N_f$\cite{ssz} if one
 considers the high density QCD with large $\mu$.

In our world with $N_c=3$ the $\eta'$ meson is not a light 
particle, although it may become lighter  for sufficiently 
hot hadronic matter as a result of partial restoration of $U(1)_A$ 
symmetry around the QCD phase transition, as argued in
\cite{Shuryak:U1a,McLerran}. At the moment, we do not know what exactly 
happens with this excitation at such conditions, there is
 no theoretical control
or sufficient lattice data. However, as was argued 
in Ref.\cite{fz}, 
 the domain wall in the theory with $N_c=3$ can still be 
{\em classically  stable}. In reality (at quantum level), it is
of course only $metastable$, but with  
a life time  much longer than a typical  QCD time scale,
$\sim \Lambda_{QCD}^{-1}$. 
The argument of Ref.\cite{fz} 
is based on the observation that one should not 
naively compare $m_{\eta'}$ mass 
with a typical hadronic scale which is the same order of
  $1 GeV$. Instead, 
one should compare the vacuum energy density 
due to the gluon degrees of freedom (it can be explicitly
expressed in terms of the gluon condensate, $E = \langle \frac{b
\alpha_s}{32 \pi} G^2 \rangle \sim N_c^2$)
 with the corresponding contribution
due to the $\eta'$ excitation when the relevant dimensionless phase
 $\frac{\eta'}{f_{\eta'}} $ becomes order of one. Therefore, 
 $\frac{1}{2}m_{\eta'}^2{\eta'}^2\sim \frac{1}{2}
m_{\eta'}^2f_{\eta'}^2\sim \frac{\partial^2}{\partial \theta^2}
E\sim N_c^2/N_c^2\sim 1 \ll E$. Exactly this inequality prevents the domain walls from the   classically allowed fast
 decay as discussed in \cite{fz}. 
In what follows we assume that $\eta' $ domain wall is
 classically stable object,
and therefore, it decays through the  quantum tunneling  process 
with exponentially large lifetime\footnote{Recall familiar  soap film bubbles: those are
metastable as well, but they do exist for minutes with sizes of many
cm, which is very impressive if expressed 
in terms microscopic (atomic)  units.
} which is  longer than any
other time scales existing in the heavy ion collisions. 

The main point of this letter is the observation that 
if such domain walls indeed exist in QCD, they can be produced and studied 
in heavy ion collisions. 
At high collision energies (SPS, RHIC) the excited
matter is assumed to be produced in the QGP phase, and then cools
down, spending significant time ( $\sim 5  fm/c $)  in the so called 
{\em mixed phase}. Small bubbles made of domain walls
can be produced by thermal fluctuations at this stage: currently  
we are not able to provide any quantitative estimates of the
probability
of this to happen, as it would require
an understanding of the ``out of equilibrium" physics.
However, one may not expect  
a huge suppression for the production of a $\sim few fm$ size  
bubble due to the general arguments that all dimensional parameters
of the problem  have one and the same 
 $\sim 1 fm$ scale during this period. In fact
the domain wall tension $\sigma$\cite{fz} :
\be
\label{sigma}
\sigma\simeq \frac{\pi^2f_{\pi}\sqrt{E}}{2\sqrt{2}N_c}\simeq 
(200 MeV)^3\simeq (1 fm)^{-3}.
\ee 
is not large enough to upset energetics at $T=T_c\approx 160 \ MeV$.
Our main observation is 
that after being produced, a small ($1 fm$ in size)  bubble would
{\em grow substantially  in size}, driven by the pion pressure,
spending a significant extra time ($\sim 5 )fm/c$ as a relatively large 
macroscopic object, before releasing all its content and disappear.

How can these   large  bubbles   be observed if produced?
Our suggestion  is to monitor (on event-by-event basis) the
strength of the
 two-pion Bose-Einstein correlation function, $\lambda_*(p)$, see below,
 which it turns 
out to be very sensitive to any macroscopically large object
with long enough life time. All pions which are eventually emitted
from such an object will be completely incoherent with the rest
of pions. 

Closing the Introduction we should mention that  related 
but  different macroscopically large configurations were also
discussed in Ref. \cite{Pisarski} in the context of the decay of the
metastable vacua possibly created in heavy ion collisions. The
difference with this work is quite significant: we do not 
consider a metastable vacuum, but rather a metastable wall making a bubble.

\section{ Bubble dynamics}

Effective Lagrangian for bubble motion is
\be
\label{b1}
L={4\pi\sigma R^2(t) \over 2} \dot{R}^2(t) - 4\pi\sigma R^2(t) +
{4\pi\over 3}R^3(t)P_{\pi}
\ee
where $\sigma$ is the wall surface tension (\ref{sigma})
and $R(t)$ is bubble radii.
 The last term describes the effective pressure induced by pions, which
competes with the surface tension and try to expand the bubble.
It appears because pions are scattered back, from the domain wall
into the bubble. The probability of this to happen  $P(k_r)$, the reflection coefficient of $\pi$ meson
with momentum $k_r$ off a domain wall, enters the pion pressure
\be 
\label{b2}
P_{\pi}=\int 2 k_r v_r n_\pi(\vec k)P(k_r){d^3k \over (2\pi)^3}, 
\ee
($P(k_r)$ will be estimated in the next section.) We assume the bubble is a 
spherically symmetric object such that we keep only the  $r$ component,  
$v_r$ is the velocity of the pions; $ n_\pi(\vec k)$ is the pion
density, to be discussed below.
The equation of motion is then
\be 
\label{b3}
\ddot{R}(t)=-{\dot{R^2} \over R} - {2 \over R} + {P_{\pi} \over \sigma},
\ee
which has perfect physical meaning as the last term describes
the acceleration of the bubble's surface. Indeed, 
$P_{\pi}$ can be interpreted as the force/area applied to the bubble,
and $\sigma$ can be interpreted as mass/area of the bubble.

The bubble radius $R(t)$ is one  time-dependent variable: the
 second one is the temperature of the hadronic gas inside the 
bubble $T(t)$. (We assume that pions rescattered back from the bubble wall
are quickly equilibrated with the rest of pions inside it.)
We therefore need the second equation, which
we get simply from the energy conservation.
One can identify the following terms in the energy change of the bubble
\be 
\label{b4}
\dot{E}_{gas}=-\dot{E}_{leakage}-\dot{E}_{wall}
\ee
where 
\be 
\label{b5}
\dot{E}_{gas}={d \over dt}( c_{SB}T^4 {4\pi R^3(t) \over 3})
\ee
and
$c_{SB}=\pi^2/10$ for massless pion Bose gas\footnote{It is  $9/\pi^2$ in Boltzmann
approximation we will use for simplicity.}.
\be 
\label{b6}
\dot{E}_{leakage}=4\pi R^2 \int  k_r v_r n_\pi(\vec k)(1-P(k_r))
{d^3k \over 2(\pi)^3}
\ee
\label{b7}
\be 
\dot{E}_{wall}=8\pi \sigma R \dot{R}
\ee
After the replacement
$\<v_rk_r\>\rightarrow \frac{1}{3}\< k \>$ in eq.(\ref{b6})
( brackets
imply the averaging over ensemble with temperature $T$),
the corresponding equation finally gets   the form
\be
\label{b8}
{4 \dot{T} \over 3 T} +{\dot{R} \over R}=-{1\over 3 R}+ 
{P_{\pi}\over 2 c_{SB}T^4 R } - {2 \sigma \dot{R} \over R^2 c_{SB} T^4}
 \ee
If the rhs is set to zero, the bubble evolution gives just 
$T^4 R^3 =const$, as energy conservation would require.
The negative sign of the rhs tends to compress the bubble.

Our main observation is that after being produced, small bubble would
be forced to grow substantially, provided some conditions (to be
specified
below) are met. This expansion may happen due to two different
reasons.
 
{\bf The first}, if a bubble happen to contain a significant
portion of its volume by QGP, it will expand
simply because while it is transformed
into hadronic matter (approximated by a pion gas) it
 occupies much larger volume.
 The simplest (but naive) estimate of
such
expansion ratio $R^{exp}_1$ 
is obtained if both the QGP and the pion gas are
treated as ideal massless gases; then it is
just the ratio of degrees of freedom (DOF)
 of QGP ($c_{QGP} =47.5, N_c=N_f=3$) to that of
the pion gas ($c_{\pi}=3$). If so, 
the volume is expected to grow by an order of magnitude.
In reality, we have the resonance gas, not the pion one,
with more degrees of freedom, and partons are not really massless.
So $R^{exp}_1$  is not $c_{QGP}/c_{\pi} \simeq 16$,
but somewhat smaller.

{\bf The second} mechanism driving expansion works
 in hadronic phase. We will
show in the next section that the thermal pions are rather effectively reflected by the
domain wall, so that they are effectively trapped inside the bubble.
It leads to another expansion ratio $R^{exp}_2=s(T_c)/s(T_{min})$
where the entropy densities at two temperatures are included. The
latter one, $T_{min}$, should correspond to final
mechanically equilibrium bubble satisfying the following condition
\be p(T_min)=\sigma/R_f \ee 
The larger the final bubble radius $R_f$, the lower this pressure. The lifetime
of this equilibrium is determined either by partial leakage of pions,
or by the bubble lifetime, whatever is shorter.


\begin{figure}[ht!]
\begin{center}
     \epsfig{file=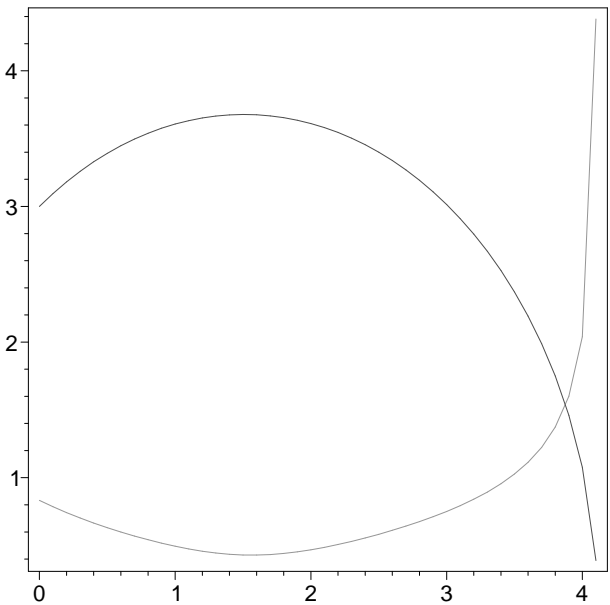, width=80mm}
\end{center}
  \caption[]
  {
   \label{bubble} An  example of the solution of equations of motion of the bubble
dynamics,
described in the text. The solid line is the dependence of the bubble
radius $R(t)$, in $fm$, versus time t,  in $fm$. We assume $R(t=0)=3 fm $
which corresponds to the initially small bubble about $1.2 fm$
expanded due to the first mechanism, $ 1.2 fm\cdot(\frac{c_{QGP}}{c_{\pi}})^{1/3}\simeq 3 fm $
up to the relatively large size $R(t=0)\simeq 3 fm $.
 The dotted line
represents the corresponding evolution of the internal temperature $T$
(in $fm^{-1}$).}
\end{figure}

 Equations (\ref{b3}) and (\ref{b8})
derived above, describe the dynamics of bubbles in this hadronic phase  and correspond to the   
second stage of expansion determined by $R^{exp}_2$. We solved equations (\ref{b3}) and (\ref{b8})
numerically to demonstrate the effect of this expansion,  see Fig.1.
Let us only comment
on T evolution. It starts at the critical value $T_c=165 \, MeV=.83 \ fm^{-1}$ and
then cools down as the bubble expands. However, the equations
indicate  a secondary heating
as the bubble walls collapses.
We will not dwell on it, but note that quite similar phenomena are known
elsewhere in physics, e.g. in the so called sonoluminescence:
so it may be not just an artifact of the equation solution.

\section{Pion scattering on a domain wall}
In the discussions presented above we introduced in our formula (\ref{b2})
one essential  parameter, the reflection probability $P(k_z)$
which is, by definition, the probability of reflection of $\pi$ meson
off a domain wall. The behavior of this
 parameter has not been specified yet, we shall estimate $P(k_z)$
now.

First of all, we start from the low energy effective Lagrangian when
$\pi$ and $\eta'$ fields are described by the unitary matrix
in the simplified version of the theory when $N_f=2, m_u=m_d$:
\be
\label{1}
U=\exp\left(i\frac{\sqrt{2}\pi^a\sigma^a}{f_{\pi}}+
i\frac{\sqrt{2}\eta'}{f_{\eta'}}\right), ~ UU^{\dagger}=1,~ , 
\ee
where $\sigma^a$ are the Pauli matrices, $\pi^a$ is the triplet, 
and $f_{\pi}\simeq f_{\eta'}\simeq 133 MeV$. In terms of $U$ the low energy
effective Lagrangian is given by\cite{HZ}:
\be
\label{2}
L=\frac{f_{\pi}^2}{8}Tr(\d_{\mu}U^{\dagger}\d_{\mu}U)+
\frac{1}{2}M Tr(U+U^{\dagger})+ \nonumber \\
E\cos(\frac{i\log Det U-\theta }{N_c}),
\ee
where all dimensional parameters in this Lagrangian are 
expressed in terms of QCD vacuum 
condensates, and are well-known:
$M=m_q |\<\bar{\psi}{\psi}\>|$;  the constant $E$ is related
to the gluon condensate $E=\<\frac{b\alpha_s}{32\pi}G^2\>$.
The first two terms describe  the standard expression for the effective chiral Lagrangian;
 the last term describes the $\eta'$ field.
This term is not uniquely fixed by the symmetry; however
the only important element   for the  following discussions
is the manifest $2\pi$ periodicity for the $\eta'$ term as well as  an 
    appearance of the scale $E$ which remains finite in the chiral limit. 
A specific $\cos$  form for this term is not essential;
however it will be used for all numerical estimates presented below.
 
As we discussed in the Introduction the theory (\ref{2}) supports
 the metastable domain walls. We refer the reader to the original paper
\cite{fz} for the discussions of  the specific properties of the domain wall;
the only what we need to know  here about the QCD
domain walls is their  following general properties:\\
a). The QCD domain wall is described by two 
dimensionless phases, $\phi_S(z)$ describing the isotopical
``singlet'', and $ \phi_T(z)$   isotopical ``triplet''
fields. These fields correspond to the dynamical $\eta'$ (singlet) and
pion $\pi^0$ (triplet) fields defined in (\ref{1}); they depend only
on one variable $z$. Both dimensionless
fields $\phi_S (z), \phi_T(z)$ interpolate between 
zero and $2\pi$ when $z$ varies from $-\infty$ to $+\infty$
and can be expressed in terms of the original 
dimensional fields $\eta', \pi^0$
as follows. 
\be
\label{3}
\phi_{S, T} = \phi_u\pm\phi_d , ~
    \eta'  = \frac{f_{\pi}}{2\sqrt{2}} \phi_S(z), ~
         \pi^0 = \frac{f_{\pi}}{2\sqrt{2}} \phi_T(z).
\ee
b). Metastable domain
walls   of a minimal energy correspond to  
  the transitions from $(\phi_u,\phi_d)|_{z=-\infty} =(0,0)$ 
to $(\phi_u,\phi_d)|_{z=+\infty}=(  2\pi, 0)$.
In the   limit $m_u=m_d$ the transition to 
$(\phi_u,\phi_d)=(0,   2\pi)$ have the same energy and
there is a degeneracy. In reality $m_d>m_u$, and the
transition to $(\phi_u,\phi_d)=(  2\pi, 0)$ is the only stable
transition. \\
c).Domain wall has a  sandwich-like  structure: $\phi_S$ substantially
varies   on the scale
$z \sim m_{\eta'}^{-1}$ while $\phi_T$ varies on  considerably larger scale
 $z \sim m_{\pi}^{-1}\gg m_{\eta'}^{-1}$ where $\phi_S$ is 
already close to its vacuum values $0,   2\pi$. Therefore, one can say that
the $\eta'$ transition is sandwiched in the pion
transition. In spite of the fact that $\phi_T$ is much wider than $\phi_S$
the main contribution to the wall tension (\ref{sigma}) comes from the $\phi_S$ transition.
Anti-soliton corresponds to the transition  from $(\phi_u,\phi_d)|_{z=-\infty} =(0,0)$  to $(\phi_u,\phi_d)|_{z=+\infty}=(  -2\pi, 0)$.

Our next step is to consider small oscillations in the pion and $\eta'$ fields  about the static  QCD domain wall solution(\ref{3}):
\be
\label{4}
\vec{\pi}(x_{\mu})\rightarrow \left( \pi_1(x_{\mu}), \pi_2(x_{\mu}),
 \frac{f_{\pi}}{2\sqrt{2}} \phi_T(z)+\pi_3(x_{\mu}) \right);
\nonumber \\
\eta'(x_{\mu})\rightarrow \frac{f_{\pi}}{2\sqrt{2}} \phi_S(z)+\eta'(x_{\mu}),
\ee
where $\phi_T(z) $ and $\phi_S(z)$  are solutions known numerically
\cite{fz}
and qualitatively described above. It is clear that a particle
scattering by a planar domain wall reduces to a one-dimensional scattering problem. To further simplify things we neglect scattering of $\eta'$ particles which can not play an important role due to its larger mass.

First, let us consider the simple case of
scattering of the neutral $\pi_0$ meson  off the domain wall.
With the plane wave ansatz, $\pi_3(x_{\mu})=\pi_3(z)
\exp(-i\omega t+ik_xx+ik_yy)$, the field equation for $\pi_3(z)$
which follows from the chiral Lagrangian (\ref{2}), can be written as
\be
\label{5}
\left(\frac{d^2}{dz^2}+k_z^2-U_{\pi^0}(z)\right)\pi_3(z)=0,~~ \nonumber
\\
k_z^2\equiv
\omega^2-k_x^2-k_y^2-m_{\pi}^2, 
\ee
where the effective potential $ U_{\pi^0}(z)$
for the problem is expressed in terms
of the domain wall profile functions $\phi_T(z) $ and $\phi_S(z)$:
\be
\label{6}
U_{\pi^0}(z)=-m_{\pi}^2\left(1-
\cos\frac{\phi_T(z)}{2}\cos\frac{\phi_S(z)}{2}\right)
\ee
Equations (\ref{5}), (\ref{6}) can not be solved analytically, 
however, a qualitative behavior of the reflection coefficient 
$P(k_z)$ can be easily understood from the following arguments.
If the particle wavelength is much smaller than the thickness of the wall,
$k_z\gg m_{\pi}$, the reflection coefficient is exponentially small 
according to the standard semiclassical arguments. In the opposite, 
the long-wavelength limit $k_z\rightarrow 0$, the wall potential (\ref{6})
can be adequately approximated by a $\delta(z)$ function,
\be
\label{7}
U_{\pi^0}(z)\simeq -\alpha\delta (z),~~~ \alpha=-\int_{-\infty}^{+\infty}
U_{\pi^0}(z)dz \sim m_{\pi}.
\ee
The scattering problem with a $\delta(z) $ function potential is easily
solved; the reflection coefficient is
\be
\label{8}
P(k_z)=\frac{\alpha^2}{\alpha^2+(2k_z)^2}.
\ee
As expected, the reflection coefficient does not depend on the 
sign of the potential $U_{\pi^0}(z)$, and $P(k_z)\rightarrow 1$ for small
 $k_z\rightarrow 0$.

Our next task is an analysis of a similar problem for the charged
$\pi^+, \pi^-$ components of $\vec{\pi}$ field.
 Some technical complications  arise here due to the fact that
the kinetic term $\sim Tr(\d_{\mu}U^{\dagger}\d_{\mu}U)$
in the low energy Lagrangian (\ref{2})
is not reduced to the  canonical kinetic terms 
$\sim 1/2(\d_{\mu}\vec{\pi})^2$ for the individual
components when the expansion in (\ref{2}) is made 
about a nontrivial classical configuration.
Instead, the kinetic term is given by
\be
\label{9}
  \frac{1}{2}Tr(\d_{\mu}U^{\dagger}\d_{\mu}U)=(\d_{\mu}\theta)^2+
(\d_{\mu}\phi)^2+\sin^2\theta(\d_{\mu}\vec{n})^2,
\ee
where, in order to simplify formula (\ref{9}),
we introduce new  variables $\theta(x),\phi(x),
\vec{n}(x)$ expressed in terms of the original fields $\vec{\pi}, 
\eta'$ (\ref{1})
as follows:
\be
\label{10}
|\vec{\pi}|\equiv \sqrt{\vec{\pi}^2};~~
\vec{n}\equiv\frac{\vec{\pi}}{|\vec{\pi}|};~~
\theta\equiv\frac{\sqrt{2}|\vec{\pi}|}{f_{\pi}};~~
\phi\equiv\frac{\sqrt{2}\eta' }{f_{\pi}}.
\ee
After some algebra we arrive to the following expression
(analogous to (\ref{5},\ref{6})) describing the scattering
of charged $\pi_1, \pi_2$ components off the wall.
\be
\label{11}
\left(\frac{d^2}{dz^2}+k_z^2-U_{\pi^{\pm}}(z)\right)\pi_i(z)=0,~~ 
 ~~i=1,2 \nonumber \\
k_z^2\equiv \omega^2-k_x^2-k_y^2-m_{\pi}^2,
\ee
where the effective potential $ U_{\pi^{\pm}}(z)$
for the charged $\pi_{1,2}$ components is expressed in terms
of the same   profile functions $\phi_T(z) $ and $\phi_S(z)$
as follows
\be
\label{12}
U_{\pi^{\pm}}(z)=m_{\pi}^2\left(
\frac{\phi_T(z)}{2}\frac{\cos\frac{\phi_S(z)}{2}}{\sin\frac{\phi_T(z)}{2}}
-1 \right)+\delta U_{\pi^{\pm}}(z),
\ee
where $\delta U_{\pi^{\pm}}(z)$ is 
a quite complicated operator defined in the Appendix 
  and  numerically will be ignored
for the qualitative discussions which follow.
Note, that in all formulae presented above the profile functions
$\phi_S(z), \phi_T(z)$ describing the wall are defined 
in the region $ -\infty < z< 0$ when $0<\phi_S(z), \phi_T(z)< \pi$.
For the positive $ z$, by symmetry,  one should replace $\phi_{S, T}\rightarrow 2\pi-\phi_{S, T}$ as explained in\cite{fz}. 
Qualitative properties of the potential $U_{\pi^{\pm}}(z)$
are the same as $U_{\pi^{0}}(z)$ discussed earlier (\ref{6}), namely,
a long-wavelength particle is reflected from the wall
with a very high probability, while in the short-wavelength limit
 the reflection probability is nearly zero. The reflection coefficient
is slightly different for the charged and neutral pions; 
however, in what
 follows we neglect that difference and approximate the 
reflection coefficient as follows
\be
\label{13}
P(k_z)\simeq e^{-\frac{k_z}{k_0}}, ~~k_0\simeq m_{\pi}.
\ee
This is our final expression which was used in the previous 
section, see eqs.(\ref{b2}), (\ref{b6}), for the numerical estimates presented on Fig1.

\section{Experimental observable}
Our main idea of the experimental observation of bubbles
 is to make use of the intensity interferometry
of pions, due to their Bose-Einstein statistics. This method, 
also known as HBT (Hanbury-Brown-Twiss) interferometry, has been
originally introduced for measuring the angular diameters 
of  stars. For pions it has been first observed in $\bar p p$
annihilation,
and explained in the famous paper by Goldhaber et al\cite{GGLP}.
Since early 70's it has been used to extract source sizes of 
hadronic fireballs, see e.g. one of the early papers on the subject by 
one of us
 \cite{Shu_HBT}.
 Recently it has been argued \cite{DimaHBT} that one can  use
pion interferometry as a sensitive tool to detect
possible increase of the $\eta'$ production in heavy 
ion collisions. We now extended the same reasoning  
 for the observation
of the bubble production.

 If a bubble is produced, it exists for some lifetime
and then decay. It can either happen due to (i) puncture of the wall, or
(ii) simple contraction, as discussed above. Both scenarios lead to
similar observable signature.

In the case of puncture, the bubble
walls decays into its underlying fields, the $\eta'$. Since it
happens after a freeze-out time for most pions
not belonging to the bubble,
most of these $\eta'$ would
not be re-absorbed, and  decay normally, most often into 5 pions.
 One may argue 
(direct following ref. \cite{DimaHBT})   that
there exist an important observable distinction between those pions
and the rest of them produced from the fireball. Large lifetime of
$\eta',\eta$ make their products $incoherent$ to other pions, reducing
the HBT peak in two-pion spectra. It happens,
because the inverse of the $\eta'$ lifetime is much smaller than
experimental energy resolution of the detectors. 

The second (and, as our estimates suggest, see Section II, more probable )
mechanism for the bubble decay is described above:
it is due to a eventual collapse of
the bubble due to pion leakage. In this case bubble surface contracts
and large population of $\eta'$ is not expected. However 
  this processes   takes long  enough time,
$\sim 5$ fm/c, so the bubble itself plays the role of long-lived object.
Again, the inverse of this time is likely to be
 below the detector resolution.
As a result, most of the  pions trapped in 
these  quasi-stationary bubbles and released later    become
incoherent with others pions. For this reason the
pions from bubbles lead to the same 
effect of not producing  a HBT peak in two-pion spectra.

The strength of the HBT correlation is characterized by the effective
intercept parameter $\lambda_*( p)$ defined as follows\cite{DimaHBT}
\be
\label{14}
\lambda_*(p)=\left[\frac{N_{direct}(p)}{ N_{direct}(p)+N_{delayed}(p)  }\right]^2 ,
\ee
where $ N_{direct} (N_{delayed}(p))$ is the one- particle- invariant momentum
distribution of the ``core'' direct  ( ``halo'' or delayed  ) pions.
As was discussed in\cite{DimaHBT}, a substantial increase in the  $\eta'$
production 
will result in appearance of a hole {\em in the low
$p_t$ region} of the effective intercept parameter $\lambda_*(p)$
centered around $p_t\simeq 138 MeV$ which represents the average $p_t$ 
of the pions coming from $\eta'$ decay.
 If a bubble is punctured,
we also expect the same signature, due to
an increase   number of $\eta'$ to be produced
with low $p_t$.

However, if the second mechanism of decay
(slow inflation and deflation of the bubble)    is prevailed,
the effect  of decreasing $\lambda_*(p)$ is different.
This is because
the emission of pions from the bubble happen through the walls, and
low energy pions are expected to be nearly completely trapped
(note, the reflection coefficient is close to one for the cold pions
and zero for the hot ones, see previous Section III).
So we expect to see a decreasing $\lambda_*(p)$
at larger p instead.

To make a numerical estimate of the effect, we note that 
 the parameter $\lambda_*$ equals to $ 1 $ for
completely coherent pions and reduced to about $0.5$
 in usual experimental conditions\footnote{There are delayed
pions due to 
``natural cocktail'' of long lived resonances.}. 
If bubbles are produced,  the intercept $\lambda_*$  would 
be additionally reduced by the factor $(1-f_{bubble})^2$
as follows from (\ref{14}). Here 
$f_{bubble}$ is the fraction of pions coming from the bubbles.
this parameter can be easily estimated 
as follows. The
bubble energy   is order of $E_{bubble}\simeq 4\pi R^2\sigma
\sim 60 GeV$, where we use $ R\sim 5 fm$ and $\sigma\sim  1 fm^{-3}$
(\ref{sigma}). If all the energy accumulated in the wall of the bubble
will go to the production of the $\eta'$ mesons 
(which will result in additional $ \sim 30 \eta'$ mesons per event )
one should expect a $100$ or so of ``incoherent'' pions  
to be   produced from the bubble. 
In  the second scenario
(inflation/deflation) for bubble decay the effect
would be  proportional to $R^3$ rather than $R^2$ and because all the pions
from the bubble incoherent,
therefore their number could be similar or even larger.

Naively this number represents
relatively small fraction of the total number of pions in each given 
event.
However, the bubble is local in rapidity.
Even at mid-rapidity at RHIC, with total pion multiplicity
about 1000 per unit rapidity, the products of ``naturally occurred''
long lived resonances make about 300 of those.  Additional 100
incoherent  pions    
  expected from the bubble makes it 400, with
the average $\lambda$ changing from $.7^2=0.5$ to $.6^2=0.36$, not
   a non-negligible effect.

Therefore, we propose to look at the
event-by-event fluctuations of parameter
$\lambda_{*}$, hunting for the tail of the distribution toward
its values $smaller$ then the average, preferably at low $p_t$.  
An unusually long  tail may indicate the bubble formation.
The thermodynamical fluctuations in the
particle composition are expected to be very small $O(1/\sqrt{N})$,
effect, obviously unable to produce any long tail by itself.

\section*{ Acknowledgments}
We wish to thank  Misha Stephanov for very 
useful  discussions during the initial stage
of this project. We are both thankful to the CERN Theoretical Physics Division 
for its invitation to visit CERN,
as well as RIKEN/BNL research center,  where part of this work has been
done.
This work is partially supported by the US-DOE grant No. DE-FG02-88ER40388.
AZ is supported in part by the National Science and Engineering
Research Council of Canada. 
\section*{Appendix A.}
Here is the definition of $\delta U_{\pi^{\pm}}(z)$ which appears in eq.(\ref{12})
in terms of the profile functions $\phi_S(z), \phi_T(z)$:
\be
\label{a1}
\delta U_{\pi^{\pm}}(z)\pi_i= \\
\left(
\frac{\phi_T(z)}{2\sin\frac{\phi_T(z)}{2}} \right)^2
\left[\phi_T\frac{d}{dz}\left( \frac{f}{\phi_T}\right)\frac{d}{dz}-
\frac{d}{dz} \left( \frac{f}{\phi_T}\frac{d\phi_T}{dz}  \right)                           \right]\pi_i \nonumber
\ee
with $f(z)$ defined as
\be
f\equiv 1- \left(
\frac{2}{\phi_T}\sin\frac{\phi_T(z)}{2}\right)^2  
\ee

\end{narrowtext}
\end{document}